\documentstyle[A4,12pt]{article}
\title{\Large \bf One-loop effective action\\
                  for Einstein gravity in\\
                  special background gauge}
\author{{\bf P.M.~Lavrov}\thanks{E-mail lavrov@tspi.tomsk.su}\\
\it Department of Mathematical Analysis,\\ Tomsk State Pedagogical
Institute,\\ 634041, Tomsk, Russia\\ \and {\bf A.A. Reshetnyak}\\ \it
Quantum Field Theory Department, Tomsk State University,\\ 634010,
    Tomsk, Russia} \date{}
\begin{document}

\maketitle

\begin{abstract}
 The one-loop effective action for Einstein gravity in a special one-parameter
 background gauge is calculated up to first order in a gauge parameter. It is
 shown that the effective action does not depend upon the gauge parameter on
 shell.
\end{abstract}

\large
\section{Introduction}
\hspace*{\parindent}
The field models are recently formulated, as a rule, in the form of gauge
theories (electrodynamics, chromodynamics, (super)gravity, (super)string,
etc.). It is well known, since quantization of such theories involves
introducing the gauge, that the Green's functions, possessing the whole
information about quantum properties of the theory, depend on choice of the
gauge (see for example [1]).
On the other hand, physical values (in particular, the S matrix) must not
depend on choice of the gauge. This fact implies that gauge dependence in
gauge theories is a special issue [2,3].
The most detailed study of the problem in question for the case of general
gauge theories with arbitrary gauges in the framework of standard Lagrangian
BRST quantization [4] is given in ref.[5]. Meanwhile, the corresponding
generalization to the case of the extended BRST quantization method [6]
is presented in ref.[7].

The study of gauge dependence for concrete field models is currently popular
[8-14], but the results presented are not always correct (see for example
refs.[15-17]). Our attention to the problem in question is due to ref.[17],
which presents calculation of the one-loop effective action within a special
class of one-parameter background gauges. In ref.[17] it is stated that the
one-loop effective action depends manifestly on choice of the gauge on shell.
This result is in contradiction with general assertions of papers [3,5].

In this connection, the present paper deals with calculation of the one-loop
effective action for Einstein gravity within the class of gauges suggested in
ref.[17]. Discrepancy with the result given in ref.[17] is found along with
the source bringing it about. The effective action is shown to
depend on choice of the gauge in a manner, being in accordance with the
statements of refs.[3,5].

In this paper we use the condensed notations suggested by De Witt [18]. The
Grassmann parity of a quantity $G$ is denoted $\varepsilon(G)$. Derivatives
with respect to fields $A^\imath$ are always understood as right and those
with respect to sources $J_\imath$ as left. For derivatives with respect to
$A^\imath$ we use the special notation $F,_\imath (A)\equiv {{\delta
F(A)}/{\delta A^\imath}}$.
In what follows, we use the terminology, now becoming generally accepted as
regards gauge field theories.
\section{Gauge dependence of effective action in Einstein gravity}
\hspace*{\parindent}
Let us consider the Einstein theory of gravity described by the classical
action
\begin {eqnarray}
{\cal S}(\bar{g}_{\mu\nu}) = - {1\over k}\int d^4x \sqrt {-\bar{g}}\bar{R}
\end {eqnarray}
%(1)
%
($g_{\mu\nu} = diag(-,+,+,+), R^\mu _{\:\nu\alpha\beta} =
\partial_\alpha \Gamma^\mu _{\nu\beta} - ..., R_{\alpha\beta} = R^\mu
_{\:\alpha\mu\beta}, R =
R^{\alpha\beta}g_{\alpha\beta}$), $k$ is the gravitational constant.

The action (1) is invariant under the general coordinate transformations:
\begin {eqnarray}
\delta \bar {g}_{\mu\nu} = \bigg(\bar{g}_{\mu\sigma}\nabla_\nu(\bar{g}) +
\bar{g}_{\nu\sigma}\nabla_\mu(\bar{g})\bigg)\eta^\sigma \equiv {\vec{H}_{\mu\nu
\sigma}(\bar{g})\eta^\sigma}
\end {eqnarray}
%(2)
(with an arbitrary vector $\eta^\sigma$)
\begin {eqnarray}
    {\cal S},_\imath(\bar {g}){\cal R}_{\alpha}^{\imath}(\bar {g})\equiv
    \frac {\delta {\cal S}(\bar {g})}{\delta {\bar g}_{\mu
\nu}}{\vec{H}}_{\mu\nu\sigma}(\bar{g})\equiv 0\;,
\end {eqnarray}
%(3)
here we use the condenced notations ${\cal S},_\imath(\bar {g})$ and ${\cal R}_
{\alpha}^{\imath}(\bar {g})$ for the classical equations of motion and the
generators of the gauge transformations respectively. In (3) we have also
introduced, for the sake of convenience, the following condensed notations:
\begin {eqnarray}
\imath = (\mu, \nu, x),\quad \alpha = (\sigma, y),
\end {eqnarray}
%(4)
%
where $\mu, \nu, \sigma,$ are the Lorentz indices,
$x, y, z$ are the space-time coordinates of the Riemann manifold.

The algebra of the local generators ${\vec{H}}_{\mu\nu\sigma}(\bar{g})$ is
closed, with the structural coeffitients not dependnig upon the fields
$\bar{g}_{\mu\nu}$ and having the form
\begin {eqnarray}
 f^\lambda_{\rho\sigma}(x,y,z)
 = \delta^\lambda_{\sigma}\delta(x-y)\partial_{\rho}\delta(x-z) -
\delta^\lambda_{\rho}\delta(x-z)\partial_\sigma\delta(x-y)\;.
\end {eqnarray}
%(5)

Since the Faddeev-Popov rules could be applied to the theory in question,
the nonrenormalizable generating functionals of the Green's functions
$Z(J)$ and the vertex functions $\Gamma ({\bar g})$ are given, with
allowance made for the condensed notations (4), in the form
\begin {eqnarray}
Z(J)&=& \int D{\bar g} \exp \bigg\lbrace{\imath\over \hbar}\bigg(S_{\psi}({\bar
g}) +
J_{\imath}{\bar g}^{\imath}\bigg)\bigg\rbrace\;,
\nonumber
\\
\Gamma({\bar g})&=& {\hbar\over \imath} {\rm ln}Z(J) - J_\imath {\bar
g}^\imath\;,\quad
{\bar g}^\imath = {\hbar\over \imath} {{\delta {\rm ln}Z(J)}\over {\delta
J_{\imath}}}\;,
\end{eqnarray}
%(6)
where $S_{\psi}({\bar g})$ is the quantum action constructed by the rules:
\begin{eqnarray}
& &S_{\psi}({\bar g}) = {\cal S}({\bar g}) - {1\over 2}\chi_\alpha({\bar
g})g^{\alpha\beta}\chi_\beta ({\bar g})
- \imath \hbar {\rm Tr\; ln} M({\bar g})\;,\nonumber
\\& &
M_{\alpha\beta}({\bar g}) = \chi_{\alpha,\imath}({\bar g}){\cal
R}^{\imath}_{\beta}({\bar g})\;.
\end{eqnarray}
%(7)
Here $\chi_\alpha({\bar g})$ is a gauge function supposed to be linear in the
fields ${\bar g}_{\mu\nu}$, while $\hbar$ is the Plank constant.

In the framework of the background-field method $\bar {g}_{\mu\nu}$ can be
represented in the form:
\begin {eqnarray}
\bar {g}_{\mu\nu} = g_{\mu\nu} + \sqrt k h_{\mu\nu}\;,
\end{eqnarray}
%(8)
%
where $g_{\mu\nu}$ is the background part of the complete field $\bar {g}_
{\mu\nu}$, satisfying the classical equations of motion, and $h_{\mu\nu}$ is
the quantum field.

By virtue of decomposition (8), the gauge transformations can be written
in the form
\begin {eqnarray}
&& \qquad\delta g_{\mu\nu}=0 \nonumber
\\
&& \sqrt k \delta h_{\mu\nu} = \bigg(\bar{g}_{\mu\sigma}\nabla_\nu(\bar{g}) +
\bar{g}_{\nu\sigma}\nabla_\mu(\bar{g})\bigg)\eta^\sigma \equiv {\vec{H}_{\mu\nu
\sigma}(\bar{g})\eta^\sigma}
\end {eqnarray}
%(9)
%

Now choose for the action (1) the gauge condition in the form of a special
one-parameter background gauge [17]
\begin{eqnarray}
& &\chi^\rho(g,h,\zeta)=\bigg\lbrace{1\over 2}\bigg[g^{\rho\tau}\nabla^{\sigma}
+ g^{\rho\sigma}\nabla^{\tau} - {1\over
2}g^{\sigma\tau}\nabla^{\rho}\bigg] \nonumber
\\
& &\qquad\qquad\qquad\qquad +
\zeta{\cdot}k{\cdot}R^{\tau\omega\sigma\rho}(g)\nabla_{\omega}\bigg\rbrace
h_{\tau\sigma} \; ,
\end{eqnarray}
%(10)
%
where $\zeta$ is the gauge parameter.

Let us now introduce the operator $L^{\rho\sigma,\mu\nu}(\zeta)$ necessary
for the calculation of the one-loop counterterms to the effective action.
It is defined by the part of the complete quantum action $S_{\psi}$
quadratic in the fields $h_{\tau\sigma}$ and minimal when $\zeta=0$:
\begin {eqnarray}
& &L^{\rho\sigma,\mu\nu}(0) = {\delta \over {\delta
h_{\rho\sigma}}}{\delta_l \over {\delta h_{\mu\nu}}}
\bigg\lbrace{\cal S}(\bar{g}) \nonumber
\\
& & - {1\over 2}\int d^4 x \sqrt {-g}
\chi^\rho(g,h,0)g_{\rho\sigma}\chi^\sigma(g,h,0)\bigg\rbrace_{\mid h=0}
\nonumber
\\
& & = \sqrt {-g}C^{\rho\sigma,\lambda\delta}\bigg\lbrace \Box
\delta^{\mu\nu}_{\lambda\delta} + P^{\mu\nu}_{\lambda\delta}\bigg\rbrace\;,
\end {eqnarray}
%(11)
%
where
\begin{eqnarray}
& &{\delta^{\mu\nu}_{\lambda\delta} =
\delta^\mu_{(\lambda}\delta^\nu_{\delta)}} \;,\nonumber
\\
& &{C^{\rho\sigma,\lambda\delta} = {1\over
4}\bigg(g^{\rho\lambda}g^{\sigma\delta} +
g^{\rho\delta}g^{\sigma\lambda} - g^{\rho\sigma}g^{\lambda\delta}\bigg)}\;,
\nonumber
\\
& &{P^{\mu\nu}_{\lambda\delta} = 2R_{\lambda}^{(\mu}{}_{\delta}{}^{\;\nu)} +
2\delta^{(\mu}_{(\lambda}R^{\nu)}_{\delta)} -
g^{\mu\nu}R_{\lambda\delta} - g_{\lambda\delta}R^{\mu\nu} -
R \delta^{\mu\nu}_{\lambda\delta}+} \nonumber
\\
& &\qquad\qquad+ {1\over 2} g_{\lambda\delta}g^{\mu\nu}R \;,
\end{eqnarray}
%\nonumber %(12)
%\end{document}%
the symbol ${\delta_l}\over {\delta h_{\mu\nu}}$ denotes the left derivative
with respect to the field $h_{\mu\nu}$, whereas the indices in brackets
imply symmetrization with a factor $\frac {1}{2}$. The Green's
functions of the gauge and ghost fields $G_{\mu\nu,\lambda\delta}(\zeta)$
and $Q^\sigma_\rho(\zeta)$ are defined when $\zeta = 0$ by the relations:
\begin {eqnarray}
& &L^{\rho\tau,\mu\nu}(0)G_{\mu\nu,\lambda\sigma}(0) =
-\delta^{\rho\tau}_{\lambda\sigma}\;,\nonumber
\\ && \nonumber \\
& &\bigg(\vec{H}_{\mu\nu\rho}(g,h){{\delta\chi^\sigma(g,h,0)}\over {\delta
h_{\mu\nu}}}\bigg)\cdot Q^\tau_\sigma (0) = \delta^\tau_\rho \; .
\end {eqnarray}
%(13)
%
Here the operator $L^{\rho\tau,\mu\nu}$ is written without $\sqrt {-g}$
and the Faddeev-Popov matrix $M_{\alpha\beta}$ in (7) has the form
\begin {eqnarray}
M_\rho^\sigma(g,h,\zeta) = \vec{H}_{\mu\nu\rho}(g,h){{\delta
\chi^\sigma(g,h,\zeta)}\over {\delta h_{\mu\nu}}}\;.
\end {eqnarray}
%(14)
%
One-loop effective action of the theory is given by:
\begin {eqnarray}
\imath \Gamma_1(\zeta) = -{1\over 2} {\rm Tr\; ln} L^{\rho\tau,\mu\nu}(\zeta)
+ {\rm Tr\; ln} M_\rho^\sigma(\zeta)\;.
\end{eqnarray}
%(15)
%
Differential consequence of the identities (3) and the relations (13) lead to
the one-loop Ward identity for the Green's functions (the condensed notations
are used)[19]:
\begin{eqnarray}
{{\delta\chi_\alpha(\zeta)}\over {\delta h^n}}G^{nm}(\zeta) =
Q^\beta_\alpha(\zeta){\cal R}^m_\beta - {\cal S},_\imath {{\delta {\cal
R}^\imath_\beta}\over {\delta h^n}} G^{nm}(\zeta)Q^\beta_\alpha(\zeta)\;.
\end{eqnarray}
%(16)
%
 From the identities (16) there follows the representation for the one-loop
effective action with an accuracy up to the first order in the gauge
parameter:
\begin {eqnarray}
& &\imath\Gamma_1(\zeta) = \imath\Gamma_1(0) + \zeta\cdot{\cal
S},_\imath {{\delta {\cal R}^\imath_\alpha}\over {\delta
h^n}} G^{nm}(0)Q^\alpha_\beta(0)\bigg({d\over {d\zeta}}
{{\delta\chi^\beta (\zeta)}\over {\delta h^m}}\bigg) \nonumber
\\
& &\qquad\qquad\qquad + O(\zeta^2)\; .
\end {eqnarray}
%(17)
%
Note that calculation of the counterterms for divergent structures in (17)
involves gauge invariant regularization for the Einstein gravity (namely,
dimensional). Absence of anomalies for the general coordinates invariance is
also taken into account.

For calculation of a divergences in (17) we applied the Barvinsky-
Vilkovisky diagrammatic technique in the dimensional regularization scheme
[19] (assuming $\delta(0) = 0$) of the Schwinger proper-time integration
method.

All diagrams the representation (17) for $\Gamma_1(\zeta)$ contains are
finite with the background dimensionalities $O({1\over {l^n}}), n>4$. There
are the following background dimensionalities of values in the theory
$[{\cal S},_\imath] = [R^\mu_{\;\nu\alpha\beta}] =
[R_{\alpha\beta}] = [R] = [{1\over k}] = O({1\over {l^2}})$. Calculation of
divergences in (17) gives a quadratically divergent diagram and two
logarithmically divergent diagrams:
\begin {eqnarray}
& &K_{1\mid div} = - \zeta\cdot k\int
\Gamma^{(\rho}_{\mu\alpha}(\nabla){\cal L}^{\mu\sigma)}\nabla_\tau
R^{\lambda\tau\delta\alpha} C^{-1}_{\rho\sigma,\lambda\delta} \nonumber \\
& &\qquad{\hat 1\over {\Box^2}}\delta(z-y)dz_{\mid y=z,div},\nonumber \\
& &I_{1\mid div} = - \zeta\cdot k\int {\cal L}^{\mu(\sigma}
R^{\lambda\tau\delta\gamma}R^\alpha_\gamma \nabla_\tau
\Gamma^{\rho)}_{\mu\alpha}(\nabla) C^{-1}_{\rho\sigma,\lambda\delta} \nonumber
\\
& &\qquad{\hat 1\over {\Box^3}}\delta(z-y)dz_{\mid y=z,div},\nonumber \\
& &I_{2\mid div} = - \zeta\cdot k \int {\cal
L}^{\mu(\sigma}R^{\lambda\tau\delta\alpha}P_{\rho\sigma,\lambda\delta}
\nabla_\tau \Gamma^{\rho)}_{\mu\alpha}(\nabla) \nonumber \\
& &\qquad{\hat 1\over {\Box^3}}\delta(z-y) dz_{\mid y=z,div},
\end {eqnarray}
\nonumber %
%\end{document}%
%(18)
where ${\cal L}^{\mu\sigma} = - \bigg(R^{\mu\sigma} - {1\over 2}
g^{\mu\sigma}R\bigg)$
is the classical extremal,
\begin{eqnarray}
& &\Gamma^\rho_{\mu\alpha}(\nabla) = \delta^\rho_\mu \nabla_\alpha -
2\delta^\rho_\alpha \nabla_\mu,
\nonumber
\\& &
C^{-1}_{\rho\sigma,\lambda\delta} = \bigg(g_{\rho\lambda}g_{\sigma\delta}
+ g_{\rho\delta}g_{\sigma\lambda} -
g_{\rho\sigma}g_{\lambda\delta}\bigg),
\nonumber
\end{eqnarray}
and the operator $P_{\rho\sigma,\lambda\delta}$ is defined in (12).

The relations (18) are reduced to a table of the universal functional traces
[19]. We only use the following ones:
\begin {eqnarray}
& &\nabla_\mu\nabla_\nu {\hat 1\over {\Box^2}}\delta(x-y)_{\mid
y=x,div} = {{\imath {\rm ln} L^2}\over {16\pi^2}}\sqrt {-g}\bigg\lbrace {1\over
6}\bigg(R_{\mu\nu} - {1\over 2} g_{\mu\nu}R\bigg)\hat 1 \nonumber \\
& &\qquad + {1\over 2}\hat R_{\mu\nu}\bigg\rbrace, \nonumber \\
& &\nabla_{\mu_1}...\nabla_{\mu_{2n-3}} {\hat 1\over {\Box
^n}}\delta(x-y)_{\mid y=x,div} = 0 ,\qquad n\geq 2, \nonumber \\
& &\nabla_\mu\nabla_\nu {\hat 1\over{\Box^3}}\delta(x-y)_{\mid y=x,div} =
{{\imath {\rm ln} L^2}\over {64\pi^2}}\sqrt {-g} g_{\mu\nu} \hat 1.
\end {eqnarray}
\nonumber %(19)
Here ${\hat R}_{\mu\nu}B = [\nabla_\mu, \nabla_\nu]B,\; L^2 $ is the parameter
of effective cutoff.

In view of (17), the resultant form for the divergent part of one-loop
effective action
\begin {eqnarray}
& &\Gamma_{1,div}(\zeta) = \Gamma_{1,div}(0) + K_{1\mid div} +
I_{1\mid div} + I_{2\mid div} =
\nonumber
\\
& & = {{\imath {\rm ln} L^2}\over {16\pi^2}}\int d^4
x\sqrt {-g}\bigg\lbrace {53\over 15} \bigg(R^2_{\mu\nu\alpha\beta}-
4R^2_{\mu\nu}+R^2\bigg) + {21\over 10} R^2_{\mu\nu} +
\nonumber
\\
& & + {1\over 20} R^2 + \zeta\cdot
k\bigg[-6R_\tau^{\;\sigma}
R^{\lambda\tau\delta\alpha}R_{(\lambda\alpha\delta)\sigma} +
{29\over 2}R^{\rho\sigma}\bigg(R_\rho^{\;\tau}{}_{\;\sigma}{}^{\;\alpha}
R_{\alpha\tau}
\nonumber
 \\
& & +R^\tau_{\;\sigma}R_{\rho\tau}\bigg) +
3RR^{\lambda\sigma\delta\alpha}R_{(\lambda\sigma\delta)\alpha} - {67\over 4}
R R^2_{\mu\nu} + {15\over 8} R^3 \bigg]\bigg\rbrace
\nonumber
\\
& &\qquad\qquad + O(\zeta^2)
\end{eqnarray}
%\nonumber %(20)
%\end{document} %
(calculation of $\Gamma_{1,div}(0)$ was given, for instance, in [19]) depends
upon the gauge parameter $\zeta$ off-shell only. This result, being the
consequence of general theorem for dependence of effective action upon the
gauge in gauge theories, is not unexpected.
\section{Effective action in general gauge theories}
\hspace*{\parindent}
Let us consider the gauge theory of fields $A^\imath(\varepsilon(A^\imath)
\equiv \varepsilon_\imath)$
with the classical action ${\cal S}(A)$ invariant under the gauge
transformations of the form: $\delta A^\imath = {\cal
R}_{\alpha}^{\imath}(A){\xi}^\alpha$ ($\xi^\alpha$ are
arbitrary functions, ${\varepsilon (\xi^\alpha )} \equiv {\varepsilon_\alpha
}$),
where ${\cal R}_{\alpha}^{\imath}(A)$ are generators of gauge transformations
assumed to obey the usual requirements of irreducibility and completeness.

%%%%%%%%%%%%%%%%%%%%%%%%%%%%%%%%%%%%%%%%%%%%%%%%%%%%%%%%%%%%%%%%%%%%%%%%%%%%%
As mentioned above, the most detailed and complete investigation of
gauge dependence in general gauge theories in the Lagrangian formulation of
standard BRST quantization [4] was given in paper [5]. From the results
[5] obtained in the framework of the standard assumptions (a gauge invariant
regularization, absence of anomalies) we only borrow those to relate
immediately to the problem in question, i.e. the one of gauge dependence
of Green's functions.
%%%%%%%%%%%%%%%%%%%%%%%%%%%%%%%%%%%%%%%%%%%%%%%%%%%%%%%%%%%%%%%%%%%%%%%%%%%%%
In [5] it is proved that both nonrenormalizable and renormalizable generating
functionals of vertex functions do not depend upon the gauge on their
extremals (in particular, the renormalizable physical S matrix does not
depend upon the gauge). This statement is valid for arbitrary gauge theories
(when the algebra of gauge transformations is both closed and open) in
arbitrary gauges.

In what follows it suffices for the purposes of this paper to confine
ourselves to consideration of the Yang-Mills type theories. In terms of the
generators of gauge transformations ${\cal R}^{\imath}_{\alpha}(A)$ these
theories are formulated as gauge ones, for which the Lie brackets of
generators ${\cal R}^{\imath}_{\alpha}(A)$ (commutator) have the form:
\begin {eqnarray}
 {{\cal R}^{\imath}_{\alpha,\jmath}(A){\cal R}^{\jmath}_{\beta}(A) -
  (-1)^{\varepsilon_{\alpha}\varepsilon_{\beta}}{\cal
  R}^{\imath}_{\beta,\jmath}(A){\cal R}^{\jmath}_{\alpha}(A)} =
 - {\cal R}^{\imath}_{\gamma}(A){F}^{\gamma}_{\alpha\beta},
\end {eqnarray}
%(21)
%
with the structural coeffitients $F^{\gamma}_{\alpha\beta}$ not depending
upon the fields $A^\imath$. The generators themselves form a complete and
linearly independent set. They are also linear with respect to the fields
$A^\imath$. For such theories one can specify the following result of ref.[5]
obtained for the first time in [3].

The quantum action of the theory is constructed by the Faddeev-Popov rules:
\begin{eqnarray}
& &S_{\psi}(A) = {\cal S}(A) - {1\over 2}\chi_\alpha(A)\chi^\alpha(A)
- \imath \hbar {\rm Tr\; ln} M(A),
\nonumber
\\& &
M_{\alpha\beta}(A) = \chi_{\alpha,\imath}(A){\cal
R}^{\imath}_{\beta}(A).
\end{eqnarray}
%(22)
%
Here $\chi_{\alpha}(A)$ is the gauge function. In what follows we shall
suppose it to be linear with respect to the fields $A^\imath$ (linear
gauges). The generating functionals of Green's functions $Z(J)$ and vertex
functions $\Gamma(A)$ are constructed by the following rules:
\begin {eqnarray}
Z(J)&=& \int DA \exp \bigg\lbrace{\imath\over \hbar}\bigg(S_{\psi}(A) +
J_{\imath}A^{\imath}\bigg)\bigg\rbrace\;,
\nonumber
\\
\Gamma(A)&=& {\hbar\over \imath} {\rm ln}Z(J) - J_\imath A^\imath,\quad
A^\imath = {\hbar\over \imath} {{\delta {\rm ln}Z(J)}\over {\delta
J_{\imath}}},
\end{eqnarray}
%(23)
%
where $J_\imath$ are the sources to the fields $A^\imath$.

The study of gauge dependence of $Z(J)$ and $\Gamma(A)$ is based on the fact
that any variation of gauge conditions in (23) leads to a change of both the
summand fixing the gauge ${1\over 2}{\chi_\alpha}{\chi^\alpha}$ and the
summand containing the Faddeev-Popov matrix $M_{\alpha\beta}$. Both variations
can be compensated by the corresponding gauge transformation of the fields
$A^\imath$, which can be considered as the change of variables in the
functional integral (23) with the Berezenian equal to 1. Thus, one can
obtain an equation, determining gauge dependence for $\Gamma(A)$.
%%%%%%%%%%%%%%%%%%%%%%%%%%%%%%%%%%%%%%%%%%%%%%%%%%%%%%%%%%%%%%%%%%%%%%%%%%%%%
Analysis of this equation implies [3] for the completely renormalizable
quantum action \[S_{\psi_{R}}(A) = S_\psi(A) -\sum\nolimits_{n=1}^\infty
{\Gamma_{n,div}},\] where $\Gamma_{n,div}$ is the divergent part of $n$-loop
approximation for $\Gamma(A)$, which can be written in the form:
%%%%%%%%%%%%%%%%%%%%%%%%%%%%%%%%%%%%%%%%%%%%%%%%%%%%%%%%%%%%%%%%%%%%%%%%%%%%%
%
\begin {eqnarray}
S_{\psi_R}(\lbrace\vartheta\rbrace,A) = \hat{S}_{\psi_R}(A'), \quad
A' = A'(\lbrace\vartheta\rbrace,A)
\end {eqnarray}
%(24)
%
($\lbrace\vartheta\rbrace$ is the set of all gauge parameters in
$\chi_{\alpha}(A)$). Here all dependence upon the gauge is contained in the
variables of gauge invariant functional $\hat{S}_{\psi_R}(A')$. In turn, for
renormalizable generating functional of vertex functions $\Gamma_R$ one can
establish
(in the class of linear gauges!) the following representation:
\begin {eqnarray}
\Gamma_R(\lbrace\vartheta\rbrace,A) = \hat{\Gamma}_R(A') - {1\over 2}
\chi_\alpha \chi^\alpha,\quad A' = A'(\lbrace\vartheta\rbrace,A),
\end {eqnarray}
%(25)
%
where (excepting the gauge condition) all dependence upon the gauge is
contained
in the variables of gauge invariant functional
$\hat{\Gamma}_R(A') =
\tilde{\Gamma}_R(\lbrace\vartheta\rbrace,A)$. The explicit form of
representation (25) enables one to draw a conclusion that the generating
functional $\tilde{\Gamma}_R$ on its extremals
\begin {eqnarray}
{{\delta\tilde{\Gamma}_R}\over {\delta A^\imath}} = 0
\end {eqnarray}
%(26)
%
does not depend upon the gauge. Indeed, variation of $\tilde{\Gamma}_R$ is
written by the relation for variation of gauge condition with respect to the
one of parameters $\vartheta$:
\begin {eqnarray}
{\delta_{\vartheta}\tilde{\Gamma}_R} = {{\delta\hat{\Gamma}_R}\over {\delta
A'^\imath}}
{{\partial A'^\imath}\over {\partial\vartheta }} \delta\vartheta
\end {eqnarray}
%(27)
%
In view of nondegeneracy of the parametrization $A' =
A'(\lbrace\vartheta\rbrace,A)$
one obtains simultaneously with (26)
${{\delta\hat{\Gamma}_R}\over {\delta A'^\imath}} = 0$. Consequently
$\tilde{\Gamma}_R$ depends upon $\lbrace\vartheta\rbrace$ off shell only.

As regards the example of Einstein gravity considered above the general
relations of this section assume the following form. The condensed notation of
indices are described by (4), the fields $A^\imath$ corresponding to the metric
tensor ${\bar g}_{\mu\nu}(x)$. Meanwhile the gauge algebra structural
coeffitientse $F^\gamma_{\alpha\beta}$ in (21) are assotiated with the
functions $f^\lambda_{\rho\sigma}(x,y,z)$ in (5) and
$\varepsilon_\imath=\varepsilon_\alpha=0$.

Concluding, note that the resultant form for the divergent part of one-loop
effective action (20) does not coincide with the result of paper [17], where
the following term is present:
\begin {eqnarray}
\zeta\cdot k\cdot
R_{\mu\nu\lambda\sigma}R^{\lambda\sigma\delta\alpha}R^{\mu\nu}_{\;\;\delta\alpha}.
\end {eqnarray}
%(28)
The reason for this structure to appear is wrong usage of the relation (15).
In [17] the contributions of ghost and gauges fields to $\Gamma_{1,div}$ have
not been calculated correctly, and therefore the term (28) remains in
$\Gamma_1(\zeta)$.
\paragraph{Acknowledgements\\}                                              

The authors are grateful to I.V. Tyutin and P.Yu. Moshin for useful
discussions. The work is supported in part by the Russian
Foundation for Fundamental Research, project N 94-02-03234 and by
International Science Foundation, grant N RI 1000.

\end{document}